\documentclass[aps,prl,twocolumn,preprintnumbers,superscriptaddress]{revtex4}

\usepackage{color}
\usepackage{multirow}

\usepackage[normalem]{ulem}

\usepackage{amsmath}
\usepackage{enumerate}
\usepackage{amsfonts}
\usepackage{epsfig}

\newcommand{\be}{\begin{equation}}
\newcommand{\ee}{\end{equation}}
\newcommand{\bea}{\begin{eqnarray}}
\newcommand{\eea}{\end{eqnarray}}

\def\le{\left}
\def\ri{\right}

\def\di{\lambda}

\begin{document}
\title{Tales of Lifshitz Tails II}
\title{Lifshitz Tails of Disordered Scale Invariant Theories}
\title{Lifshitz Tails of Scale Invariant Theories with Electric Impurities}
\title{Lifshitz Tails of Scale-Invariant Theories with Electric Impurities}

\preprint{MIT-CTP-4386}

\author{Sho Yaida}
\affiliation{Center for Theoretical Physics, Massachusetts Institute of Technology,
Cambridge, MA 02139, USA
%{\tt yaida@mit.edu}
}

%\date{July, 2012}

\begin{abstract}
We study scale-invariant systems in the presence of Gaussian quenched electric disorder, focusing on the tails of the energy spectra induced by disorder.
For relevant disorder we derive asymptotic expressions for the densities of unit-charged states in the tails, positing the existence of saddle points in appropriate disorder integrals.
The resultant scalings are dictated by spatial dimensions and dynamical exponents of the systems.
\end{abstract}
\maketitle

\newpage

%\section{Enlarging Impure Thoughts}\label{introduction}
The study of Lifshitz tails -- the tails of energy spectra induced by quenched impurities -- is an ancient subject~\cite{Lifshitz}.
For noninteracting particles obeying the Schr{\" o}dinger equation with a random potential, there exist various systematic methods for obtaining an asymptotic expression for the density of states deep in the tail~\cite{HL1, ZL, HL2, Cardy, HS, LW, ICofLT}.
The goal of the present paper is to broaden our perspective on Lifshitz tails by exploring a class of systems whose low-energy excitations are governed by scale-invariant theories.

The organization of the paper is as follows.
We first set up a scale-invariant theory with quenched electric disorder.
After defining the disorder-averaged density of unite-charged states, we argue that there exists a family of saddle points in the appropriate disorder integral.
The asymptotic expression for the density of unit-charged states is then derived for large negative energy.
We conclude with assertions to be ideally proven in order to rigorously establish the existence of the saddle points postulated herein.

%\section{Scale Invariant Theory}\label{SIT}
A clean scale-invariant system possesses a dilatation operator $\hat{D}$ along with a time-translation operator $\hat{H}_0$ and space-translation operators $\hat{P}_i$ for $i=1, ..., d$. These operators obey
\be
[\hat{H}_0, \hat{P}_i]=0,\ [\hat{P}_i, \hat{P}_j]=0,\ [\hat{D}, \hat{P}_i]=i\hat{P}_i,
\ee
and
\be
[\hat{D}, \hat{H}_0]=i z \hat{H}_0
\ee
where $z$ is a dynamical exponent.
We suppose that the system has a conserved current with a local ``charge" density operator $\hat{J}^{t}({\bf x})=e^{-i\hat{\bf P}\cdot{\bf x}}\hat{J}^{t}({\bf 0})e^{+i\hat{\bf P}\cdot{\bf x}}$ obeying $[\hat{J}^{t}({\bf x}), \hat{J}^{t}({\bf y})]=0$ and
\be
[\hat{D}, \hat{J}^{t}({\bf 0})]=i d \hat{J}^{t}({\bf 0}).
\ee
A charge number operator $\hat{Q}\equiv\int d{\bf x}\hat{J}^{t}({\bf x})$ in particular satisfies $[\hat{H}_0, \hat{Q}]=[\hat{D}, \hat{Q}]=0$.
We also suppose that there is a local operator $\hat{\cal O}^{\dagger}_{\rm pro}({\bf x})=e^{-i\hat{\bf P}\cdot{\bf x}}\hat{\cal O}^{\dagger}_{\rm pro}({\bf 0})e^{+i\hat{\bf P}\cdot{\bf x}}$ with scaling dimension $\Delta_{\rm pro}$ and unit, minimal, charge number $q_{\rm unit}$. In other words,
\be
[\hat{D}, \hat{\cal O}^{\dagger}_{\rm pro}({\bf 0})]=i\Delta_{\rm pro}\hat{\cal O}^{\dagger}_{\rm pro}({\bf 0})
\ee
and
\be
[\hat{Q}, \hat{\cal O}^{\dagger}_{\rm pro}({\bf 0})]=q_{\rm unit}\hat{\cal O}^{\dagger}_{\rm pro}({\bf 0}).
\ee 
We set $\hbar=1$ and $q_{\rm unit}\equiv1$ henceforth.

%\section{Gaussian Quenched Electric Disorder}\label{QED}
Let us now sprinkle ``electric" impurities into the clean system, deforming the Hamiltonian to
\be
\hat{H}_{V_{\rm random}}=\hat{H}_0+\int d{\bf x} V_{\rm random}({\bf x})\hat{J}^{t}({\bf x})
\ee
where for concreteness we suppose that a random potential $V_{\rm random}({\bf x})$ obeys Gaussian statistics.
An intensive observable $O$, when scanned over a macroscopic sample, typically self-averages and we can legitimately estimate it by means of a disorder integral as~\cite{footSA}
\be\label{SA}
\le[O\ri]_{\rm d.a.}=\int\frac{\le[{\cal D}V\ri]}{{\cal N}_{\sharp}} e^{-\frac{1}{2\gamma}\int d{\bf x}V^2({\bf x})}O^V.
\ee
Here, $O^V$ is the value of the observable in the system governed by $\hat{H}_{V}\equiv\hat{H}_0+\int d{\bf x} V({\bf x})\hat{J}^{t}({\bf x})$ with a square-integrable potential $V({\bf x})$, ${\cal N}_{\sharp}\equiv\int\le[{\cal D}V\ri] e^{-\frac{1}{2\gamma}\int d{\bf x}V^2({\bf x})}$ is the normalization constant for the disorder integral, and $\gamma$ characterizes the strength of the disorder.
For each realization of $V({\bf x})$ we label eigenstates as
\be
\hat{Q}|Q; n\rangle_V=Q|Q; n\rangle_V
\ee
and
\be
\hat{H}_V|Q; n\rangle_V=E^V_{Q;n}|Q; n\rangle_V.
\ee

%\section{Density of Unit-Charged States}\label{DoS}
We probe the dirty system by injecting a unit-charged excitation through $\hat{\cal O}^{\dagger}_{\rm pro}$ and observing how it propagates.
Specifically we look at a local density of unit-charged states~\cite{footG} defined via
\be
\rho_{\hat{\cal O}^{\dagger}_{\rm pro}}^{V}(E, {\bf x})\equiv-\frac{1}{\pi}{\rm Im}\le\{G_{\hat{\cal O}^{\dagger}_{\rm pro}}^{V}({\bf x}, {\bf x}; E)\ri\}
\ee
to be disorder-averaged where
\bea
G_{\hat{\cal O}^{\dagger}_{\rm pro}}^{V}({\bf x}, {\bf y}; E)&\equiv&-i\int dt e^{iEt}\theta(t)\\
&&\times\ _V\langle0;0|\hat{\cal O}^{V}_{\rm pro}(t, {\bf x})\hat{\cal O}^{V\ \dagger}_{\rm pro}(0, {\bf y})|0;0\rangle_V\nonumber
\eea
with $\hat{\cal O}^{V\ \dagger}_{\rm pro}(t, {\bf x})\equiv e^{+i\hat{H}_{V} t}\hat{\cal O}^{\dagger}_{\rm pro}({\bf x})e^{-i\hat{H}_{V} t}$.
Here, $|0; 0\rangle_{V}$ denotes a state of the lowest energy among states with zero total charge for a given $V({\bf x})$~\cite{footCE}.
When applied to noninteracting systems, this definition reproduces the standard density of states for a particle excited by $\hat{\cal O}^{\dagger}_{\rm pro}$.

%\section{Spectral Representation}
The density of unit-charged states $\rho_{\hat{\cal O}^{\dagger}_{\rm pro}}^{V}(E, {\bf x})$ defined above has the spectral representation~\cite{Negele}
\be\label{SR}
\sum_n| _V\langle Q=1; n|\hat{\cal O}^{\dagger}_{\rm pro}({\bf x})|0; 0\rangle_V|^2\delta(E-E^V_{1;n}+E^V_{0;0}).
\ee
Contributions for negative energy $E$, if any, come from bound states with $E^V_{1;n}-E^V_{0;0}=E<0$ and a nonzero overlap $ _V\langle Q=1; n|\hat{\cal O}^{\dagger}_{\rm pro}({\bf x})|0; 0\rangle_V\ne0$.
When disorder-averaged, they give rise to a smooth Lifshitz tail.
We are interested in the asymptotic behavior of $\le[\rho_{\hat{\cal O}^{\dagger}_{\rm pro}}(E)\ri]_{\rm d.a.}$ in the limit of large negative energy $E$.
%Note that in the absence of disorder $\rho_{\hat{\cal O}^{\dagger}_{\rm pro}}(E)$ vanishes for negative $E$, rightly demanding that the energy spectrum is bounded below for a fixed number of charges (dilatation argument).

%\section{Localons}\label{postulates}
At this point we make two postulates, both of which can be rigorously established for a noninteracting scale-invariant theory with $z=2$~\cite{ICofLT}.
First we assume that for any square-integrable potential $V({\bf x})\ne0$, when $d<2z$, there exists a state (or states) of lowest energy $E^V_{1;0}$ among states with a unit charge excited by $\hat{\cal O}^{\dagger}_{\rm pro}$.
%For nonrelativistic ($z=2$) and relativistic ($z=1$) noninteracting systems, one can rigorously establish the existence of a unique normalizable ground state for $d\leq4$ in the presence of a square-integrable potential~\cite{rigorousQM}.
Then, as emphasized in~\cite{ICofLT}, the game is to seek a localizing potential which minimizes the cost $\int d{\bf x}V^2\le({\bf x}\ri)$ while still holding a bound state with $E^V_{1;0}-E^V_{0;0}=E$ for a fixed negative energy $E$ so that it contributes to $\le[\rho_{\hat{\cal O}^{\dagger}_{\rm pro}}(E)\ri]_{\rm d.a.}$.
And here comes our second postulate:
for $d<2z$, there exists a family of square-integrable potentials $V^{E}_{\rm saddle}({\bf x})$
which minimizes the cost among all the square-integrable potentials with $E^V_{1;0}-E^V_{0;0}=E$.
Generically we expect that the competition between the cost, preferring narrower and shallower potential wells, and the demand for trapping a bound state with a fixed negative energy settles into such minimizers.
In the saddle-point approximation
\be
\le[\rho_{\hat{\cal O}^{\dagger}_{\rm pro}}(E)\ri]_{\rm d.a.}\sim{\rm exp}\le[-\frac{1}{2\gamma}\int d{\bf x}\le\{V^{E}_{\rm saddle}({\bf x})\ri\}^2\ri]
\ee
then yields the leading exponential factor.

%\section{Lifshitz Tails}\label{LT}
Armed with the two postulates, we can now obtain the asymptotic expression for the density of unit-charged states in the tail via simple dimensional analysis.
Let us be as pedantic as possible, however.
First we can use commutation relations to show that
\be
e^{-i\di \hat{D}}\hat{H}_Ve^{+i\di \hat{D}}=e^{z\di}\hat{H}_{V^{(\di)}}
\ee
with
\be
V^{(\di)}({\bf x})=e^{-z\di}V(e^{-\di}{\bf x}),
\ee
from which we deduce that
\be
e^{-i\di \hat{D}}|Q; n\rangle_V=|Q; n\rangle_{V^{(\di)}}
\ee
with the scaling relation of the spectra
\be
E^{V^{(\di)}}_{Q;n}=e^{-z\di} E^V_{Q;n}.
\ee
Combined with the scaling relation of the cost
\be\label{costscaling}
\int d{\bf x}\le\{V^{(\di)}({\bf x})\ri\}^2=e^{(d-2z)\di}\le[\int d{\bf x}\le\{V({\bf x})\ri\}^2\ri],
\ee
we conclude that for $E^{(\di)}=e^{-z\di} E$
\be\label{saddlescaling}
V^{E^{(\di)}}_{\rm saddle}({\bf x})=\le\{V^{E}_{\rm saddle}({\bf x})\ri\}^{(\di)}=e^{-z\di} V^{E}_{\rm saddle}(e^{-\di}{\bf x}).
\ee
From Eqs.~(\ref{costscaling}) and~(\ref{saddlescaling}) it then follows that
\be
\frac{1}{2\gamma}\int d{\bf x}\le\{V^E_{\rm saddle}({\bf x})\ri\}^2=\frac{a_0}{g(E)}
\ee
with the dimensionless constant $a_0$ and the dimensionless disorder coupling
\be
g(E)=\gamma(-E)^{\frac{d}{z}-2}.
\ee
Thus in the saddle-point approximation
\be\label{tail}
\le[\rho_{\hat{\cal O}^{\dagger}_{\rm pro}}(E)\ri]_{\rm d.a.}\sim e^{-\frac{a_0}{g(E)}}
\ee
for $d<2z$.
This expression is valid in the regime $E\ll -\gamma^{\frac{z}{2z-d}}$ where the disorder coupling $g(E)$ is small, akin to the dilute instanton gas limit.
We see that the asymptotic scaling of the Lifshitz tail is dictated by the spatial dimension and the dynamical exponent, ordaining the dispersion relation of the low-energy excitations.
The scaling dimension $\Delta_{\rm pro}$ enters only into the subleading prefactor.
%Restriction to the operator with a definite dimension: spectral weight scales only by power-law, much weaker than the exponential law.
%Evaluation of the subleading prefactor would involve zero-mode analysis and Gaussian integrations of nonzero modes (and renormalization thereof), which is beyond the scope of the present paper.

Our result conforms with the existing result~\cite{HL1, ZL, HL2, Cardy, HS, LW, ICofLT} for a noninteracting scale-invariant system with $z=2$.
It is also in accord with the Harris criterion~\cite{Harris} which stipulates that the disorder is relevant for $d<2z$.
We can apply our formula to any scale-invariant systems, interacting or not, such as quantum critical materials~\cite{Sachdev} and relativistic systems with massless charged excitations.

%\section{Daydreaming, Impurely}\label{conclusion}
We end with two assertions for unitary scale-invariant theories which, if proven, would ensure the validity of Eq.(\ref{tail}).
We state them more strongly here than we did in the body of the paper, respecting close parallels with statements proven for noninteracting particles obeying the Schr{\" o}dinger equation (for which the second statement becomes equivalent to the instanton problem extensively analyzed in~\cite{CGM, BGK, BL}).
\begin{enumerate}
\item
For $d\leq 2z$, for any square-integrable potential $V({\bf x})$, $\hat{H}_V$ admits a normalizable state (or states) of lowest energy for $Q=0$ and for $Q=q_{\rm unit}\equiv1$~\cite{footO}.

\item
For $d<2z$, for a fixed negative energy $E$, there exists a family of monotone spherically symmetric potentials vanishing at infinity, labeled by the translational collective coordinates, which minimizes the cost $\int d{\bf x}V^2\le({\bf x}\ri)$ among all the square-integrable potentials with $E^V_{1;0}-E^V_{0;0}=E$.

\end{enumerate}
%Any models of interacting scale invariant systems for which one can corroborate these assertions would be appreciated.
On top of proving these statements, it would be valuable to see how far one can generalize the result presented herein: in reality there are quenched disorders other than electric impurities, disorder distributions need not be Gaussian, and we can probe dirty systems through operators with nonminimal charges~\cite{footU}.

%\begin{acknowledgments}
The author thanks Allan~W.~Adams, Patrick~A.~Lee, Hong~Liu, John~A.~McGreevy, and Michael~C.~Mulligan for generously sparing their time for discussions.
He also thanks Shamit Kachru, John~A.~McGreevy, and Michael~C.~Mulligan for many constructive comments on the manuscript.
He especially wishes to express his gratitude to Allan~W.~Adams for collaboration on related questions, which evolved into a seed for the present paper.
He is supported by a JSPS Postdoctoral Fellowship for Research Abroad.
%\end{acknowledgments}

\appendix

\section{APPENDIX}

In this appendix we derive coupled equations which determine saddle points of the disorder integral.
Recall that we are seeking for minima of the cost $\int d{\bf x}V^2\le({\bf x}\ri)$ with the constraint $E^V_{1;0}-E^V_{0;0}=E$.
Through the introduction of a Lagrange multiplier $\lambda_0$, the problem becomes equivalent to the minimization of
\bea
I\le[V\le({\bf x}\ri), \lambda_0\ri]\equiv+\frac{1}{2}\int d{\bf x}V^2\le({\bf x}\ri)+\lambda_0(E^V_{1;0}-E^V_{0;0}-E).
\eea
Extremizing it with respect to $\lambda_0$ reproduces the constraint
\be\label{E1}
E^V_{1;0}-E^V_{0;0}=E
\ee
while extremizing it with respect to $V({\bf x})$ yields
\be\label{E2}
V\le({\bf x}\ri)=-\lambda_0\le[_V\langle 1; 0|\hat{J}^{t}({\bf x})|1; 0\rangle_V-\ _V\langle 0; 0|\hat{J}^{t}({\bf x})|0; 0\rangle_V\ri].
\ee
Here,
\be\label{E3}
\hat{H}_V|0; 0\rangle_V=E^V_{0;0}|0; 0\rangle_V
\ee
and
\be\label{E4}
\hat{H}_V|1; 0\rangle_V=E^V_{1;0}|1; 0\rangle_V,
\ee
and we used the Hellmann-Feynman relation~\cite{Hellmann, Feynman}
\be
\frac{\delta}{\delta V({\bf x})}E^V_{Q;0}=\ _V\langle Q; 0|\hat{J}^{t}({\bf x})|Q; 0\rangle_V.
\ee
Solving the coupled equations (\ref{E1}), (\ref{E2}), (\ref{E3}), and (\ref{E4}), we obtain saddle points of the disorder integral in question.
We then further seek for a family of saddle points which minimizes the cost among all the saddles.


\begin{thebibliography}{11}

\bibitem{Lifshitz}
  I.~M.~Lifshitz,
  Adv.\ Phys.\  {\bf 13}, 483 (1964).

\bibitem{HL1}
  B.~I.~Halperin and M.~Lax,
  %``Impurity-Band Tails in the High-Density Limit. I. Minimum Counting Methods,"
  Phys.\ Rev. {\bf 148}, 722 (1966).

\bibitem{ZL}
  J.~Zittartz and J.~S.~Langer,
  %``Theory of Bound States in a Random Potential,"
  Phys.\ Rev. {\bf 148}, 741 (1966).

\bibitem{HL2}
  B.~I.~Halperin and M.~Lax,
  %``Impurity-Band Tails in the High-Density Limit. II. Higher Order Corrections,"
  Phys.\ Rev. {\bf 153}, 802 (1967).

\bibitem{Cardy}
  J.~L.~Cardy,
  %``Electron localisation in disordered systems and classical solutions in Ginzburg-Landau field theory,"
  J.\ Phys.\ C:\ Solid\ State\ Phys.\ {\bf 11}, L321 (1978).

\bibitem{HS} 
  A.~Houghton and L.~Schafer,
  %``High Order Behavior Of Zero Component Field Theories Without The N ---> 0 Limit,''
  J.\ Phys.\ A A {\bf 12}, 1309 (1979).

\bibitem{LW}
%(general formula, applied to different classes of disorder)
  J.~M.~Luttinger and R.~Waxler,
  %``Low energy density of states in several disordered systems,"
  Ann.\ Phys.\ {\bf 175}, 319 (1987).

\bibitem{ICofLT} 
  S.~Yaida,
  %``Instanton Calculus of Lifshitz Tails,''
  arXiv:1205.0005 [cond-mat.dis-nn].

\bibitem{footSA}
For a detailed discussion of when and why such disorder-averaged quantities give extremely accurate estimates of observable quantities for a macroscopic sample with quenched disorder, see~\cite{SpinGlasses}. We presume that the local density of unit-charged states defined herein is not too sensitive to boundary conditions
%in particular to which valleys of nearly degenerate ground states one belongs to,
and thus self-averages.

\bibitem{SpinGlasses}
  K.~Binder and A.~P.~Young,
  %``Spin glasses: Experimental facts, theoretical concepts, and open
  %questions,''
  Rev.\ Mod.\ Phys.\  {\bf 58}, 801 (1986).

\bibitem{footG}
It is constructed deliberately so that, when spectrally decomposed, it receives contributions only from states with charge $q_{\rm unit}=1$ and never from states with charge $-1$ [cf. Eq.(\ref{SR})].
In the time domain it equals the time-ordered Green function for $t>0$ and vanishes for $t<0$.
Hence it can be related to the spectral density function and thus can be measured experimentally in principle.

\bibitem{footCE}
%We work with the canonical ensemble at zero temperature with zero charge, as opposed to the grand canonical ensemble with zero average chemical potential.
When there are several conserved currents, we set all the charges to zero.

\bibitem{Negele}
  For a review, see
  J.~W.~Negele and H.~Orland,
  {\it Quantum Many-Particle Systems}
  (Addison-Wesley, Reading, MA 1988).
  
%\bibitem{rigorousQM}
%  E.~H.~Lieb and M.~Loss,
%  {\it Analysis}, 2nd ed. Graduate Studies in Mathematics, Vol. 14, 2nd ed.
%  (American Mathematical Society, Providence, RI 2001).

\bibitem{Harris} 
  A.~B.~Harris,
  %``Effect of random defects on the critical behaviour of Ising models,"
  J.\ Phys.\ C\ {\bf 7},\ 1671 (1974).

\bibitem{Sachdev}
For a review, see
  S.~Sachdev,
  {\it Quantum Phase Transitions}
  (Cambridge University Press, Cambridge, 1999).

\bibitem{CGM} 
  S.~R.~Coleman, V.~Glaser and A.~Martin,
  %``Action Minima Among Solutions to a Class of Euclidean Scalar Field Equations,''
  Commun.\ Math.\ Phys.\  {\bf 58}, 211 (1978).

\bibitem{BGK}
  H.~Berestycki, T.~Gallou{\"e}t and O.~Kavian,
  %``Equations de champs scalaires euclidiens nonlinŽaires dans le plan (French) [Nonlinear euclidean scalar field equations in the plane],"
  C.\ R.\ Acad.\ Sci.\ Paris\ {\bf 297}, 307 (1983).

\bibitem{BL}
  H.~Brezis and E.~H.~Lieb,
  %``Minimum action solutions of some vector field equations,''
  Commun.\ Math.\ Phys.\  {\bf 96}, 97 (1984).

\bibitem{footO}
We work within the sector of unit-charged states which have nonzero overlaps $ _{V}\langle Q=1; n|\hat{\cal O}^{\dagger}_{\rm pro}({\bf x})|0; 0\rangle_{V}\ne0$.

\bibitem{footU}
For a probe operator with nonminimal charge $|q_{\rm pro}|>|q_{\rm unit}|$, an excitation may fragment into several unit-charged constituents which subsequently may prefer to be trapped in separate localized wells rather than in one deep well.

\bibitem{Hellmann}
  H.~Hellmann,
  %``"
  Einf\"{u}hrung in die Quantenchemie (Franz Deuticke, Leipzig, 1937).

\bibitem{Feynman}
  R.~P.~Feynman,
  %``Forces in Molecules,''
  Phys.\ Rev.\  {\bf 56}, 340 (1939).

\end{thebibliography}
\end{document}